\documentclass[a4paper,11pt]{article}
\pdfoutput=1 

\usepackage{jcappub} 

\usepackage[T1]{fontenc} 
\newcommand{\be}{\begin{equation}}
\newcommand{\ee}{\end{equation}}
\newcommand{\dd}{{\rm d}}
\newcommand{\HH}{{\cal H}}
\newcommand{\DT}{{\Delta T}}
\newcommand{\nn}{{\nonumber}}

\usepackage{color}

\title{Cosmological Signatures of Anisotropic Spatial Curvature}

\author[a]{Thiago S. Pereira,}
\author[b]{Guillermo A. Mena Marug\'an}
\author[c]{and Saulo Carneiro}

\affiliation[a]{Departamento de F\'isica, Universidade Estadual de Londrina, 86057-970, Londrina -- PR, Brazil}
\affiliation[b]{Instituto de Estructura de la Materia, IEM-CSIC, Serrano 121, 28006, Madrid, Spain}
\affiliation[c]{Instituto de F\'isica, Universidade Federal da Bahia, 40210-340, Salvador -- BA, 
Brazil}

\emailAdd{tspereira@uel.br}
\emailAdd{mena@iem.cfmac.csic.es}
\emailAdd{saulo.carneiro@pq.cnpq.br}

\abstract{If one is willing to give up the cherished hypothesis of spatial isotropy, many 
interesting cosmological models can be developed beyond the simple anisotropically expanding 
scenarios. One interesting possibility is presented by {\it shear-free} models in which the 
anisotropy emerges at the level of the curvature of the homogeneous spatial sections, whereas the 
expansion is dictated by a single scale factor. We show that such models represent viable 
alternatives to describe the large-scale structure of the inflationary universe, leading to a 
kinematically equivalent Sachs-Wolfe effect. Through the definition of a complete set of spatial 
eigenfunctions we compute the two-point correlation function of scalar perturbations in these 
models. In addition, we show how such scenarios would modify the spectrum of the CMB assuming that 
the observations take place in a small patch of a universe with anisotropic curvature.}

\begin{document}
\maketitle
\flushbottom
\section{Introduction}
\label{sec:intro}

The combination of observational data from modern galaxy surveys with those from the Cosmic 
Microwave Background (CMB) radiation strongly suggests that, on scales around 100~Mpc and above, 
the spatial distribution of the cosmic web is homogeneous and isotropic. This highly symmetric 
description of the universe comprises what is known as the cosmological principle, and is 
mathematically encoded in the Friedmann-Lema\^itre-Robertson-Walker 
(FLRW) spacetime metric.

Besides being backed up by observational data, the use of the FLRW metric has a technical 
advantage. Since the only free parameter in this metric is a function of time, background dynamical 
equations become ordinary differential equations, which can be solved by means of known analytical 
or semi-analytical techniques. More important, though, is the fact that, in this description of the 
spacetime, ``pictures'' of the three-dimensional spatial section at different times are copies 
of one another, differing only by a global conformal factor. If one further expands cosmological 
observables in terms of the eigenfunctions of the spatial Laplace-Beltrami operator, the 
aforementioned simplification implies that the eigenmodes of physical quantities will not 
couple through the evolution of the universe, as far as the perturbative regime is maintained, and 
then one can trace their dynamics individually by means of linear differential equations. The 
easiness and versatility of this program has enormously boosted the field of cosmology in the past
decades, culminating with the so-called standard cosmological model (the $\Lambda$CDM model), in 
which the qualitative features of the large-scale universe can be understood by means of only six 
free parameters.

As comfortable as this description might seem, it is important to keep in mind that the cosmological 
principle constitutes a set of working hypotheses, and, in reality, a particularly critical set, 
given the intrinsic human limitations to perform observations at cosmological scales. In other 
words, one might be careful to not let arguments of simplicity prevent the discovery of new 
cosmological phenomena. 

The necessity of scrutinizing the symmetry hypothesis of the standard cosmological model 
also has the important role of testing the robustness of cosmological observations, which could
otherwise be biased by {\it a posteriori} assumptions. As a matter of fact, while the observed 
approximate isotropy of the CMB is in agreement with the cosmological principle, this does 
not mean that different symmetry principles are ruled out by it and, in fact, different 
principles are known to be compatible with the CMB observations. Indeed, in the past few years 
several authors have considered the possibility that the spectrum of CMB could be fitted by large 
void models in which we would  be located near its 
center~\cite{GarciaBellido:2008nz,Nadathur:2010zm,Biswas:2010xm,Yoo:2010qy}, in clear
contrast to the homogeneity hypothesis. The spectrum of CMB fluctuations has also been contrasted 
with spatially homogeneous but anisotropic cosmological models, either as a mean of constraining the 
impact of spatial anisotropy on CMB data~\cite{martinez1995delta,Maartens:1994qq,Maartens:1995hh}, 
or as an attempt to explain known statistical anomalies of the 
CMB~\cite{Campanelli:2006vb,Gumrukcuoglu:2006xj,Pontzen:2007ii,Rodrigues:2007ny}.
Of these two different routes, the latter is the one which we plan to follow in this work. 
That is, we assume spatial homogeneity of the universe and investigate the impact of 
spatial anisotropies on the large-angle spectrum of CMB temperature fluctuations. 

The simplest example of an anisotropic spacetime corresponds to a generalization of the FLRW metric in which each spatial direction expands according to its own scale factor. Such geometries are known as Bianchi I metrics. Given that the CMB signal points to a great degree of spatial isotropy of the early universe, Bianchi I metrics can be deployed in two different ways. For example, one could suppose that the onset of inflation has taken place in an anisotropic spacetime, and ask for its impact on primordial quantum fluctuations re-entering the horizon at the time of CMB formation~\cite{Pitrou:2008gk}. On a different context, the possibility of a late-time anisotropization of the spacetime could leave the CMB spectrum essentially unchanged, while affecting both the growth rate of  structures~\cite{Koivisto:2008ig} as well as weak-lensing observables~\cite{Pitrou:2015iya,Pereira:2015jya}.
As in the case of FLRW geometries, Bianchi I metrics also enjoy the simplicity of having unperturbed 
quantities that obey ordinary differential equations. However, the anisotropic expansion of the 
spacetime inevitably leads to a coupling of Fourier modes, which renders the use of this metric 
quite ingenious at the technical level.

However, once we are willing to abandon the isotropy hypothesis of the cosmological principle, much 
richer models can be constructed besides the Bianchi I solution~\cite{Barrow:1997sy,Barrow:1997mj,Barrow:1998ih,Barrow:2001pi,Barrow:1985tda}. In fact, it has been 
demonstrated in references~\cite{Mimoso:1993ym,Barrow:1998ih,cmm,Carneiro:2001fz} that the anisotropy of the universe 
does not need to emerge from its dynamical expansion; instead, it could result from the curvature of spatial 
sections, while having the expansion controlled by one scale factor, at the cost of having an 
imperfect fluid sourcing the anisotropic curvature. This idea has been successfully applied in the 
context of late-time anisotropies in 
references~\cite{Koivisto:2010dr,Menezes:2012kc,Miranda:2014ema}. From the phenomenological point 
of view, this possibility offers two main advantages. First, since the background metric is 
controlled by a single scale factor, the redshift of electromagnetic radiation is 
exactly isotropic, which means that the CMB remains exactly isotropic at the background level. 
Second, the fact that there is a single scale factor implies that pictures of the universe at 
different times are also copies of each other, exactly as in the FLRW description. Then, in the 
perturbative regime, eigenmodes of physical quantities will also evolve independently of one 
another, something which greatly facilitates the computation of cosmological observables. 

In this work we consider two spacetime metrics presenting isotropic expansion but anisotropic 
spatial curvature. These are the metrics from the Bianchi type III and Kantowski-Sachs families. 
Following the formalism of linear and gauge-invariant perturbations introduced in~\cite{pcm}, we 
evaluate the impact of these metrics on the large-angle spectrum of CMB fluctuations, assuming that 
the particle horizon during CMB formation is much smaller than the curvature scale of these spaces. 
After introducing an explicit set of spatial eigenfunctions in Section~\ref{eigenfunctions},
we show in Section~\ref{scalar_perturbations} that, provided that the universe has gone an early 
period of exponential expansion, the dynamics of scalar perturbations will be described by one 
single gravitational potential $\Phi$. From this result, an important corollary follows: 
given that large-angle CMB fluctuations result from the geodesics of photons in a perturbed 
spacetime, the Sachs-Wolfe effect in the considered geometries will have exactly the same functional 
form as in the FLRW case. This is demonstrated in Section~\ref{2pcf}, where, using the explicit 
eigenfunctions found in Section~\ref{eigenfunctions}, we compute the temperature two-point 
correlation function in these spaces, as well as their perturbative series around the isotropic 
two-point correlation function in the limit of large curvature radius. The effect of anisotropic 
curvature leads to off-diagonal multipolar correlations in the spectrum of the CMB which 
extends to a large multipolar range. This signature is investigated numerically in 
Section~\ref{numerics}. We conclude with some remarks and perspectives of further developments in 
Section~\ref{conclusions}.

\section{Geometry and spatial eigenfunctions}\label{eigenfunctions}

We start this section by introducing the main mathematical and technical tools which will be needed later on to compute the angular power spectrum of the CMB. 

\subsection{Anisotropy through spatial curvature}

In this work we are interested in spatially homogeneous but {\it anisotropic} solutions to Einstein equations which nonetheless display an {\it isotropic} expansion, i.e., solutions whose dynamics is described by a single scale factor. Rather, the anisotropy of these solutions arises from the 
spatial curvature of the sections of constant comoving time, curvature which is not the same in all directions. This can be achieved by considering spatial sections which are the Cartesian product of 
curved subspaces. Since one-dimensional manifolds are trivially flat, the only possibility left for 
a three-dimensional manifold to present anisotropic spatial curvature is to be the product of a 
two-dimensional and a one-dimensional manifolds. In the case of maximally symmetric two-dimensional 
manifolds, and again apart from trivial flat cases, we only have to consider the possibility of a 
two-sphere or a bi-dimensional pseudo-sphere. Therefore, the spatial topology turns out to be 
$\mathbb{S}^2\times\mathbb{R}$ or $\mathbb{H}^2\times\mathbb{R}$, respectively, accepting the topology of the real line for the third spatial dimension. The metric of the spaces that we are going to consider is thus conformally (ultra-)static and can be written as
\be
\label{metric_cartesian}
\dd s^2 = a^2(\eta)\left[-\dd\eta^2+\gamma_{ab}\dd x^a\dd x^b+\dd z^2\right]\,,
\ee
where $\{a,b\}$ represent coordinates in the two-dimensional subspace and $z$ is the coordinate of the real line. For future reference it will prove convenient to rewrite the line element in 
cylindrical coordinates
\be
\label{background_metric}
\dd s^2 = a^2(\eta)\left[-\dd\eta^2+\dd\rho^2+S^2(\rho)\dd\varphi^2+\dd z^2\right]\,,
\ee
where the function $S(\rho)$ is given by
\be
S^2(\rho)=\begin{cases}
\rho^{2}\,, & \quad (\mathbb{R}^3)\\
\sinh^2\rho\,, & \quad (\mathbb{H}^2\times\mathbb{R})\\
\sin^2\rho\,. & \quad (\mathbb{S}^2\times\mathbb{R})
\end{cases}
\ee
and where, to facilitate comparison with standard results, we have included the flat FLRW case 
($\mathbb{R}^3$) as well. Furthermore, note that while $z\in(-\infty,+\infty)$ and 
$\varphi\in[0,2\pi)$ in all three cases, the variable $\rho$ belongs to the interval $[0,+\infty)$ in the $\mathbb{R}^3$ and $\mathbb{H}^2\times\mathbb{R}$ cases, while $\rho\in[0,\pi)$ for 
$\mathbb{S}^2\times\mathbb{R}$. Moreover, note that we can always recover the flat space limit by 
taking $\rho\ll1$, which gives
\[
S^2(\rho)\sim\rho^2\,.
\]
In the cosmology literature, metrics described by either the $\mathbb{H}^2\times\mathbb{R}$ or the 
$\mathbb{S}^2\times\mathbb{R}$ spatial slices are known as Bianchi type III (BIII) and 
Kantowski-Sachs (KS) solutions, respectively.

\subsection{Spatial eigenfunctions}

In order to compute correlation functions between physical observables (for instance, the 
gravitational potential), it is very convenient to deal with their spatial dependence by expanding 
them in modes that decouple in the dynamical field equations. These field equations are typically of 
a generalized Klein-Gordon type. Hence, for the case of scalar observables, we will first need to 
explicitly find and normalize (with respect to the inner product given by the volume element of the 
spatial sections) a complete set of scalar modes which are solutions of the eigenvalue problem
\be
\nabla^2\mathcal{Q} = -q^2\mathcal{Q}\,,\qquad q\in\mathbb{R}\label{eigenproblem}
\ee
where $\mathcal{Q}$ is a scalar function, and
\be
\nabla^2 \equiv \frac{1}{\sqrt{h}}\partial_i\left(\sqrt{h}\,\partial^i\right)\,,\quad
h={\rm det}(h_{ij}),
\ee
is the Laplace-Beltrami operator in a three-dimensional Riemannian space with metric $h_{ij}$. The eigenfunctions in the three cases considered here can be found by means of a simple separation of variables
\[
\mathcal{Q}(\rho,\varphi,z) = f(\rho)g(\varphi)h(z)\,.
\]
Below, we present these normalized eigenfunctions.

\subsubsection{Flat FLRW}

In the flat FLRW case the spatial metric is $h_{ij} = \mathrm{diag}(1,\rho^2,1)$. Using separation 
of variables, the normalized solutions to the eigenvalue 
problem~\eqref{eigenproblem} are easily found to be
\be
\label{FLeigenfunc}
Z_{\omega mk}(\mathbf{x})\equiv \omega^{1/2}J_{m}(\omega\rho)\frac{e^{im\varphi}}{\sqrt{2\pi}}\frac{e^{ikz}}{\sqrt{2\pi}}\,,
\ee
where the eigenvalues are
\[
\omega\in\mathbb{R}^{+},\quad m\in\mathbb{Z}\,,\quad k\in\mathbb{R}\,,
\]
and obey the dispersion relation
\be
\label{dispersionFL}
q^2=\omega^{2}+k^{2}\,.
\ee
The normalization constant is such that
\be
\label{orthog_fl}
\int\sqrt{h}\,\dd^{3}\mathbf{x}\, Z_{\omega mk}(\mathbf{x})Z_{\omega'm'k'}^{*}(\mathbf{x})=\delta_{mm'}\delta(\omega-\omega')\delta(k-k')\,.
\ee
Any scalar function ${\cal Q}(\mathbf{x})$ can be expanded in terms of this basis as
\be
\label{eq:expansion_FL}
\mathcal{Q}(\mathbf{x})=\sum_{m=-\infty}^{\infty}\int_{0}^{\infty}\dd\omega\int_{-\infty}^{+\infty}\dd k\,\mathcal{Q}_{m}(\omega,k)\, Z_{\omega mk}(\mathbf{x})\,,
\ee
with inverse given by
\be
\mathcal{Q}_{m}(\omega,k) = \int\sqrt{h}\,\dd^3\mathbf{x}\,\mathcal{Q}(\mathbf{x})Z^*_{\omega m k}(\mathbf{x})\,.
\ee

\subsubsection{Bianchi III}

In the BIII case the spatial metric in cylindrical coordinates is $h_{ij}=\rm{diag}(1,\sinh^2\rho,1)$. The normalized solutions in this case are 
\be
\label{B3eigenfunc}
Z_{\omega mk}(\mathbf{x}) \equiv N_{m\omega}P_{-\frac{1}{2}+i\omega}^{m}(\cosh\rho)\frac{e^{im\varphi}}{\sqrt{2\pi}}
\frac{e^{ikz}}{\sqrt{2\pi}}\, , 
\ee
where
\[
\omega\in\mathbb{R}^{+},\quad m\in\mathbb{N}\,,\quad k\in\mathbb{R}\,,
\]
and
\be
\label{dispersionB3}
q^2=\omega^{2}+k^{2}+\frac{1}{4}\,.
\ee
Note that, owing to the curvature of the spatial sections, there is a strictly positive lower bound 
to the frequency of a mode:
\be
\label{limitB3}
q^2 \geq \frac{1}{4}\,.
\ee
Modes with $0\leq q^2\leq1/4$ do not belong to the basis of square-integrable eigenfunctions of the 
Laplace-Beltrami operator, and are thus excluded from the frequency spectrum. Such frequencies lead 
to what is known as supercurvature modes~\cite{Lyth:1995cw}.

The Legendre function $P_{-\frac{1}{2}+i\omega}^{m}(\cosh\rho)$ -- sometimes also called conical 
function -- is real and has the following symmetries
\be
P^{m}_{-1/2+i\omega}(x) = P^{m}_{-1/2-i\omega}(x)
= \frac{\Gamma(i\omega+m+1/2)}{\Gamma(i\omega-m+1/2)}P^{-m}_{-1/2+i\omega}(x)\,,\qquad (x\geq1)\,.
\ee
The (real) normalization constant $N_{m\omega}$ is given by 
\be
\label{norm_b3}
N_{m\omega} =\sqrt{(-1)^{m}\omega\tanh\pi\omega\frac{\Gamma(i\omega-m+1/2)}{\Gamma(i\omega+m+1/2)}}\,,
\ee
and is such that \cite{Bander:1965im}
\be
\label{orthog_b3}
\int\sqrt{h}\,\dd^{3}\mathbf{x}\, Z_{\omega mk}(\mathbf{x})Z_{\omega'm'k'}^{*}(\mathbf{x})=\delta_{mm'}\delta(\omega-\omega')\delta(k-k')\,.
\ee
Since these eigenfunctions form a basis for the Hilbert space of square-integrable functions with 
the volume element corresponding to $h_{ij}$, any such function $\mathcal{Q}(\mathbf{x})$ can be 
expanded as
\begin{equation}
\label{eq:expansion_b3}
\mathcal{Q}(\mathbf{x})=\sum_{m=-\infty}^{+\infty}\int_{0}^{\infty}\dd\omega\int_{-\infty}^{+\infty}\dd k\,
\mathcal{Q}_{m}(\omega,k)Z_{\omega m k}(\mathbf{x})\,,
\end{equation}
with the inverse following directly from the orthogonality relation \eqref{orthog_b3}. The 
flat-space limit is reached when $\rho\ll 1$ and the frequency is large, $\omega\gg 1$, such that 
$\omega\rho$ is kept fixed. Using the approximations
\begin{eqnarray}
&&(i\omega)^mP^{-m}_{i\omega-1/2}(\cosh\rho)\sim (-1)^mJ_{m}(\omega\rho),\\ 
&&\frac{\Gamma(i\omega+m+1/2)}{\Gamma(i\omega-m+1/2)}\sim\omega^{2m},
\end{eqnarray} 
one can check that, up to an overall and unimportant phase,
\be
Z_{\omega m k }(\mathbf{x}) \sim \omega^{1/2} J_{m}(\omega\rho)\frac{e^{im\varphi}}{\sqrt{2\pi}}\frac{e^{ikz}}{\sqrt{2\pi}}\,,
\ee
as it should be.

\subsubsection{Kantowski-Sachs}

Finally, for the KS case, the spatial metric in cylindrical coordinates is 
$h_{ij} = (1,\sin^2\rho,1)$. Since the spatial sections are just the product of the unit two-sphere 
with the real line, the eigenfunctions are simply the standard spherical harmonics multiplied by a 
one-dimensional plane wave:
\be
\label{KSeigenfunc}
Z_{\ell m k}(\mathbf{x}) \equiv 
N_{m\ell}P_{\ell}^{m}(\cos\rho)\frac{e^{im\varphi}}{\sqrt{2\pi}}\frac{e^{ikz}}{\sqrt{2\pi}} \,,
\ee
where
\[
\ell\in\mathbb{N}\,,\quad m\in\mathbb{Z}\,,\quad k\in\mathbb{R}\,,
\]
and we now have
\be
\label{dispersionKS}
q^2=\ell(\ell+1)+k^{2}\,.
\ee
Although the $\ell=0$ mode is included in the basis of square-integrable functions, this mode 
corresponds to an overall monopole, and is excluded from the definition of cosmological perturbations. Thus, the lower bound to the frequency of a mode is
\be
\label{limitKS}
q^2\geq 2\,.
\ee
The normalization constant is defined as
\be
N_{m\ell} \equiv \sqrt{\frac{(2\ell+1)(\ell-m)!}{4\pi(\ell+m)!}}
\ee
and is chosen to give
\be
\label{orthog_ks}
\int\sqrt{h}\,\dd^{3}\mathbf{x}\, Z_{\ell mk}(\mathbf{x})Z_{\ell'm'k'}^{*}(\mathbf{x})=\delta_{\ell\ell'}\delta_{mm'}\delta(k-k')\,.
\ee
Since these solutions also form a complete basis for the space of square-integrable functions, we 
can again expand any such function as
\be
\mathcal{Q}(\mathbf{x}) = \sum_{\ell=0}^\infty\sum_{m=-\ell}^\ell\int_{-\infty}^{+\infty}\dd k\,\mathcal{Q}_{\ell m}(k) 
Z_{\ell m k}(\mathbf{x})\,,
\ee
with the inverse following from \eqref{orthog_ks}. The flat-space limit may again be achieved by 
letting $\rho\ll 1$ and $\ell\gg1$, with $\ell\rho$ held constant. Using the identity 
$P^{-m}_{\ell}=(-1)^m(\ell-m)!/(\ell+m)!P^m_{\ell}$ and the approximations 
$P^{-m}_{\ell}(\cos\rho)\sim\ell^{-m} J_{m}(\ell\rho)$ and $(\ell+m)!/(\ell-m)!\sim\ell^{2m}$, we 
find
\be
\label{ZKSlim}
Z_{\ell m k }(\mathbf{x}) \sim \ell^{1/2} J_{m}(\ell\rho)\frac{e^{im\varphi}}{\sqrt{2\pi}}\frac{e^{ikz}}{\sqrt{2\pi}}\,,
\ee
again, up to a phase. 

\section{Scalar perturbations}\label{scalar_perturbations}

In this section we recall the basic formalism of linear perturbation theory in BIII and KS 
backgrounds~\cite{pcm}. Then, by analyzing the linearized Einstein equations of the perturbative 
degrees of freedom we show that, provided that the universe has gone an early period of exponential 
expansion, scalar perturbations can be parameterized by a single free function, exactly as in 
the standard FLRW case. Our main result is \eqref{final_metric}. Readers not interested in the 
details of the derivation can skip them and go directly to Section~\ref{2pcf}. Throughout this 
section we use the coordinates introduced in formula \eqref{metric_cartesian}.

\subsection{General formalism}

The theory of linear cosmological perturbations is built upon the idea that the large-scale universe 
can be described by means of small perturbations added to a background (exact) solution of Einstein 
equations. In general terms, this program can be implemented with the introduction of ten arbitrary 
functions on top of the background spacetime metric. Then, for homogeneous and isotropic spatial 
sections, these functions can be classified as either Scalars, Vectors or Tensors (SVT) by 
exploring their transformation properties under the action of the spatial symmetry group. In an 
FLRW universe this leads to the known SVT mode decomposition~\cite{Peter:1208401}. The main 
advantage of this procedure is that, in the linear regime, each kind of perturbation evolves 
independently of the others, which simplifies the analysis of their dynamical evolution. In this 
decomposition, non-trivial SVT modes exist in all FLRW cosmologies whose spatial slices are copies 
of a maximally symmetric space. In the BIII or KS spacetimes, on the other hand, spatial slices are 
given by the product of the real line with a two-dimensional maximally symmetric space. Therefore, 
in this case it seems natural to classify the perturbations modes just according to their behavior 
with respect to the symmetry group of that two-dimensional manifold~\cite{pcm}. In this sense, it is 
worth remarking  that there is no non-trivial transverse and traceless symmetric tensor in two 
dimensions. More specifically, on the two-dimensional maximally symmetric subspaces, any vector 
$V_a$ and symmetric tensor $T_{ab}$ can be decomposed as
\begin{align}
V_a & = D_aV+\bar{V}_a\,,\\
T_{ab}& = X\gamma_{ab} + D_aD_bY+ D_{(a}\bar{Z}_{b)}\,,
\end{align}
where $D_a$ is the covariant derivative compatible with the metric $\gamma_{ab}$ 
[see~\eqref{metric_cartesian}] and barred vectors are transverse, i.e., 
$D_a\bar{V}^a=0=D_a\bar{Z}^{\,a}$. Let us also notice that, similarly, no 
non-trivial transverse vector can exist in the remaining one-dimensional section of the spatial 
slices, in which only scalar modes can be defined. Actually, this decomposition of the perturbations 
is an adaptation of the more general formalism of Gerlach and Sengupta~\cite{GS}, extended beyond 
spherical symmetry, where we treat the two dimensions supplementary to those of the bi-dimensional maximally symmetric manifold not in a covariant way as a Lorentz manifold on its own, but rather described in terms of the conformal time and the $z$-coordinate introduced above.

Since the background metric of the BIII and KS models considered here have a single scale factor, 
there will be no dynamical coupling between SV modes, something which further simplifies the 
dynamical evolution of the system. As stated in the Introduction, this situation is in sharp 
contrast with perturbation theory in Bianchi I universes where, although one can define uncoupled 
SVT modes at a given initial time, the anisotropic evolution of the background metric couples those 
modes, leading to a dynamical see-saw effect~\cite{Pereira:2007yy,Pitrou:2008gk}. Once clarified 
that perturbative mode couplings will not arise in the present case, we can focus exclusively on the 
evolution of scalar perturbations, which are the most important ones for the CMB large-angle 
perturbations. The most general metric with scalar perturbations can be written
\be
\label{deltag1}
\dd s^2 = a^2(\eta)[-(1+2\Phi)\dd\eta^2-2\Pi\dd\eta\dd z + (1+2\Psi)\gamma_{ab}\dd x^a\dd x^b+
(1+2\Lambda)\dd z^2]\,,
\ee
where $\Phi$, $\Pi$, $\Psi$, and $\Lambda$ represent the four effective scalar degrees of freedom 
of the metric.\footnote{Seven scalar functions in the original parameterization minus three scalar 
gauge functions~\cite{pcm}.} 

Next, we need to introduce perturbations to the matter sector. Such perturbations can be 
parameterized in the same way as above, with the addition of extra SV degrees of freedom to the 
matter components. One important aspect of the formalism is that, since the considered models do not have a dynamical FLRW limit, the anisotropic spatial curvature needs to be sourced by an 
imperfect matter component~\cite{Mimoso:1993ym}. This can be done either by introducing a stress 
component in the energy-momentum tensor~\cite{Mimoso:1993ym} or a two-form 
field~\cite{Koivisto:2010dr}. We can also take the simpler route provided by the choice of scalar 
fields, which can be chosen to be a real massless field in the BIII 
case~\cite{cmm,Carneiro:2001fz}, or a complex massless field in the KS case~\cite{MenezesCarneiro}. 
Thus, our matter sector is composed of two fields: the inflaton, whose perturbations can be 
introduced exactly as in the FLRW case (see, for example, \cite{Mukhanov:1990me}), and the above 
anisotropic and massless scalar field $\phi(\eta,\mathbf{x})$. Their perturbations are as follows:
\begin{align}
a^2\delta T^{0}_{(\phi)0} & = \Lambda - \partial_z\delta\phi\,, \\
a^2\delta T^{0}_{(\phi)z} & = -\Pi - \delta\phi'\,, \\
a^2\delta T^{a}_{(\phi)b} & = (\Lambda - \partial_z\delta\phi)\delta^a_b\,, \\
a^2\delta T^{z}_{(\phi)a} & = \partial_a\delta\phi\,, \\
a^2\delta T^{z}_{(\phi)z} & = -\Lambda + \partial_z\delta\phi\,,
\end{align}
where the prime denotes the derivative with respect to $\eta$. As we are going to show, the 
perturbation $\delta\phi$ of the anisotropic scalar field is not dynamically important in the 
inflationary stage, and can be safely ignored. 

In conclusion, given the above perturbations to both the geometrical and the matter sectors, linear 
and gauge-invariant dynamical equations can be straightforwardly computed~\cite{pcm}.

\subsection{Inflation and metric perturbations}

We are primarily interested in computing large-angle effects ($\gtrsim1^\circ$) of the BIII and KS 
geometries on the temperature angular spectrum of the CMB. At these scales, temperature 
perturbations are mainly driven by quantum fluctuations of the early universe -- a period which, we assume, was sourced by scalar fields. Then, from the $a-b$ component of the linearized Einstein equation, and from the fact that scalar fields do not produce anisotropic stress, it follows that (see (A.5) in~\cite{pcm})
\be
-D_{(a}D_{b)}(\Lambda+\Phi) = 0\,,\qquad a\neq b\,.
\ee
The above equation translates into the following constraint:
\be
\label{constraint1}
\Lambda = -\Phi\,.
\ee
This is in close analogy to the situation of standard (FLRW) perturbation theory, where the absence of anisotropic stress is used to eliminate one scalar degree of freedom of the metric. 

Given the constraint~\eqref{constraint1}, metric scalar perturbations are still described by three 
free functions: $\Phi$, $\Psi$ and $\Pi$. Surprisingly, though, only one degree of freedom is 
relevant during inflation. In order to see that, let us consider the $z-a$ component of Einstein 
equations:
\be
\label{constraint2}
-(\partial_z\Phi+\partial_z\Psi)+\HH\Pi+\frac{1}{2}\Pi'=\delta\phi\,,
\ee
where $\HH$ is the Hubble parameter. If inflation is sourced by a slowly-rolling scalar field, 
the variable $\delta\phi$ obeys \cite{pcm}
\be
\label{eq:vartheta}
w''+\left[q^2+\frac{2+3\epsilon}{\eta^2}\right]w = 0\,,\qquad w\equiv a\delta\phi\,,
\ee
where $\epsilon \equiv 1-{\cal H}'/{\cal H}^2$ is the slow-roll parameter and 
$\eta\in(-\infty,0^{-})$ is the conformal time during inflation. Note that, since the ``mass'' term 
$(2+3\epsilon)/\eta^2$ is strictly positive, $w$ will never grow, which in turn implies that 
$\delta\phi$ will be suppressed by the expansion. Indeed, for $q^2\gg\eta^{-2}$, 
$w\sim \exp[\pm iq\eta]$ and $\delta\phi$ decays as $1/a$. For $q^2\ll\eta^{-2}$, $w\sim\exp[\pm 
i\sqrt{3\epsilon+7/4}\log(-\eta)]/(-\eta)^{1/2}$, which implies that\footnote{We recall that, 
during slow-roll inflation, $a\sim1/(-\eta)$ at dominant order in $\epsilon$.} 
$\delta\phi\sim1/a^{3/2}$. Since this result holds for both BIII and KS, we can set
\be
\label{vartheta}
\delta\phi = 0\,
\ee
with no loss of generality.

If we now rewrite \eqref{constraint2} in momentum space ($\partial_z\rightarrow ik$) and use 
\eqref{vartheta}, we find that the constraint
\be
\label{constraint3}
ik[\Phi(q,\eta)+\Psi(q,\eta)] = \frac{[a^2\Pi(q,\eta)]'}{2a^2}
\ee
must hold for {\it all values of} $k$. Moreover, from the dispersion relations~\eqref{dispersionB3} 
and~\eqref{dispersionKS} we see that $q$ and $k$ can be treated as independent variables whose 
specific values fix $\omega$ (BIII) or $\ell$ (KS). In other words, the case $k=0$ in the 
constraint~\eqref{constraint3} must hold for all possible values of $\Pi(q,\eta)$. This is only 
possible if $a^2\Pi(q,\eta)$ is just a function of $q$, that is, if $\Pi(q,\eta)=f(q)/a^2(\eta)$. We 
thus see that the variable $\Pi$ is dynamically suppressed as well, and can be ignored. Since this 
result holds for both BIII and KS geometries, we can set
\be
\Pi = 0\,,\qquad \Psi = -\Phi\,.
\ee

We have thus arrived at the important conclusion that, provided that the universe has experienced 
an early period of exponential expansion, the metric describing large-scale perturbations in both 
BIII and KS geometries depends on a single free function, and can be parameterized as
\be
\label{final_metric}
\dd s^2 = a^2(\eta)[-(1+2\Phi)\dd\eta^2+(1-2\Phi)h_{ij}\dd x^i\dd x^j]\,.
\ee
This is formally the same as the metric of adiabatic scalar perturbations in FLRW universes. 
However, our result is more general because $h_{ij}(x^k)$ can  represent the FLRW, the BIII, or the KS spatial metrics.

\section{Sachs-Wolfe effect and correlation functions}
\label{2pcf}

The main conclusion of the previous section is encoded in \eqref{final_metric}, and it implies that 
scalar perturbations over the considered manifolds with anisotropic spatial curvature are 
described by a single gravitational potential, exactly like standard perturbation theory in FLRW 
universes. Moreover, \eqref{final_metric} implies that \emph{all} dynamical equations are formally 
the same as in the standard isotropic case -- the only difference being that covariant derivatives 
are now taken with respect to the more general metric $h_{ij}$. A corollary of this result is that 
the Sachs-Wolfe (SW) effect, being purely kinematical, is given by the same known expression
\be
\label{SW-effect}
\DT(\hat{\mathbf{n}}) = \frac{1}{3}\Phi(\mathbf{x},\eta)\,,
\ee
where $\hat{\mathbf{n}}$ is the direction of detection of a photon emitted at spacetime position 
($\mathbf{x},\eta$) (i.e., the coordinates of the last-scattering surface) and, again, $\mathbf{x}$ 
are the coordinates in the more general spacetimes under discussion. In this description, the effect 
of the anisotropic spatial curvature appears, in particular, through the non-trivial spectrum of 
the Laplace-Beltrami operator, and affects the angular power spectrum of the CMB at large scales, as 
we are now going to see.

\subsection{FLRW case}

We move now to the computation of the temperature two-point correlation function (2pcf) resulting 
from the Sachs-Wolfe effect in a flat FLRW universe. While this result is well-known, this section 
is intended to introduce the main formalism, and also to facilitate comparison with new results 
which will follow below. Moreover, here we will carry out the computation using cylindrical 
coordinates, that are not the ones most commonly employed.

The temperature correlation function between two CMB photons arriving from the last scattering 
surface, considering only the SW effect, is defined by
\be
\label{temp2pcf}
\left\langle \DT(\hat{\mathbf{n}})\DT(\hat{\mathbf{n}}')\right\rangle 
\equiv C(\hat{\mathbf{n}},\hat{\mathbf{n}}') 
= \frac{1}{9}\langle\Phi(\mathbf{x})\Phi(\mathbf{x}')\rangle\,,
\ee
where, for simplicity, we have dropped the instants of emission, since these should be clear from 
the context. We now use~\eqref{FLeigenfunc} to decompose the gravitational potential 
$\Phi(\mathbf{x})$ and use the fact that, in a Gaussian and statistically isotropic universe, the 
correlation between modes takes the form
\be
\label{pspecFLRW}
\langle\Phi_{m}(\omega,k)\Phi^{*}_{m'}(\omega',k')\rangle = {\cal P}(q)\delta_{mm'}
\delta(\omega-\omega')\delta(k-k')\,.
\ee
Here ${\cal P}(q)$ is the primordial power spectrum and $q$ is defined as in \eqref{dispersionFL}. 
Employing the summation theorem for Bessel functions~\eqref{bessel_id1}, we then arrive at
\be
\label{2pcfFL}
C(\hat{\mathbf{n}},\hat{\mathbf{n}}')
=\frac{1}{(6\pi)^2}\int_0^\infty\omega\dd\omega\int_{-\infty}^{+\infty} \dd k 
\, {\cal P}(q) \, J_0(\omega\Delta\rho_0)e^{ik\Delta z} ,
\ee
where we have introduced the definitions
\be
\label{deltarhoFL}
(\Delta\rho_0)^2 \equiv \rho^2 + \rho'^2 - 2\rho\rho'\cos\Delta\varphi\,,
\quad \Delta\varphi \equiv \varphi-\varphi'\,,
\quad \Delta z \equiv z - z'\,.
\ee

As we will see, expressions~\eqref{2pcfFL} and~\eqref{deltarhoFL} play a central role in the 
derivation of results for geometries with anisotropic curvature. Just for completeness, though, let 
us note that the right hand side of~\eqref{2pcfFL} can be further simplified by the change of 
variables 
\be
\label{var1}
\omega = q\sin\psi\qquad\mbox{and}\qquad k = q\cos\psi\,, 
\ee
and the subsequent integration over the range\footnote{The range of $\psi$ is conditioned by the restriction $\omega\geq0$.} $0\leq\psi\leq\pi$. The fact that the 2pcf depends only on the angle 
between $\hat{\mathbf n}$ and $\hat{\mathbf n}'$ can be seen by introducing spherical coordinates 
through 
\be
\label{var2}
\rho^{(\prime)}=\Delta\eta\sin\theta^{(\prime)}\quad\textrm{and}\quad
z^{(\prime)}=\Delta\eta\cos\theta^{(\prime)}\,,
\ee
where $\Delta\eta$ is the conformal distance to the last scattering surface. This leads to
\be
\label{C_of_theta}
C(\hat{\mathbf{n}},\hat{\mathbf{n}}') \equiv C(\vartheta) 
= \frac{1}{18\pi^{2}}\int_{0}^{\infty}q^{2}\dd q\,{\cal P}(q)\,\frac{\sin\left[q\Delta\eta\sqrt{2-2\cos\vartheta}\right]}
{q\Delta\eta\sqrt{2-2\cos\vartheta}}\,,
\ee
where $\cos\vartheta \equiv \cos\theta\cos\theta'+\sin\theta\sin\theta'\cos\Delta\varphi$. From here, the angular power spectrum $C_\ell$ can be directly computed by means of a Legendre transform over $\vartheta$:
\be
\label{angular2pcf}
C_\ell = 2\pi \int_{-1}^1\dd(\cos\vartheta)\,P_\ell(\cos\vartheta)\,C(\vartheta)\,,
\ee
which, after performing the integral in $\cos\vartheta$ by means of the change of variable 
$2y^2=1-\cos\vartheta$, leads to the well known expression~\cite{Peter:1208401}
\be
C_{\ell} = \frac{2}{9\pi}\int_0^\infty q^3{\cal P}(q)\,j_{\ell}^2(q\Delta\eta)\;\dd\ln q\,.
\ee

\subsection{Bianchi III case}

In the BIII geometry the temperature 2pcf is still defined by~\eqref{temp2pcf}. However, the 
gravitational potential is now decomposed using the eigenfunctions~\eqref{B3eigenfunc}. The 
perturbations in this case are supposed to be Gaussian but statistically anisotropic, so that the 
most general correlation function between modes will have the form 
\be
\label{fourier2pcfB3}
\langle\Phi_m(\omega,k)\Phi^{*}_{m'}(\omega',k')\rangle
= {\cal \bar{P}}(\mathbf{q}) \delta_{mm'}\delta(\omega-\omega')\delta(k-k')\,.
\ee
Since we are working with homogeneous spaces, both Kronecker and Dirac deltas are needed to ensure 
translational invariance~\cite{Abramo:2010gk}. The power spectrum, on the other hand, is a general 
function of the vector $\mathbf{q} \equiv (\omega,m,k)$, since the underlying geometries are 
anisotropic. In principle, a rigorous quantization procedure is required to fix and normalize 
$\bar{\cal P}(\mathbf{q})$ properly. Nonetheless, note that invariance under rotations in the 
maximally symmetric bi-dimensional subspace requires a power spectrum independent of $m$. Moreover,
since we are interested in departures from the CMB features typical of an FLRW universe 
resulting from the BIII geometry, it suffices to fix $\bar{\cal P}(\mathbf{q})$ by demanding that 
the two-point correlation functions in these geometries agree in the limit in which the two points coincide (limit in which the spatial curvature is ignorable). This condition leads to
\be
\label{pspecB3}
{\cal \bar{P}}(\mathbf{q}) = \frac{{\cal P}(\tilde{q})}{\tanh\pi\omega}\,,
\ee
where ${\cal P}$ is the same function appearing in~\eqref{pspecFLRW} and where we have defined 
$\tilde{q}^2\equiv\omega^2+k^2=q^2-1/4$ in order to account for the discrepancy between the 
dispersion relation in the FLRW case and \eqref{dispersionB3}. The factor $\tanh\pi\omega$ is needed 
to ensure the correct functional dependence of the 2pcf when 
$\hat{\mathbf{n}}\rightarrow\hat{\mathbf{n}}'$, according to our arguments. Using the above 
expression and the summation theorem~\eqref{legendre_id1b} for Legendre 
polynomials, we arrive at
\be
\label{2pcfB3}
C(\hat{\mathbf{n}},\hat{\mathbf{n}}') =
\frac{1}{(6\pi)^{2}}\int_{0}^{\infty}\omega \dd\omega\int_{-\infty}^{+\infty}\dd k\,
{\cal P}(\tilde{q})P_{-\frac{1}{2}+i\omega}(\cosh\Delta\rho)e^{ik\Delta z}\,,
\ee
where
\be
\label{deltarhoB3}
\cosh\Delta\rho\equiv\cosh\rho\cosh\rho'-\sinh\rho\sinh\rho'\cos\Delta\varphi\,.
\ee

Before continuing our analysis, it is interesting to compare these formulas with their 
counterparts in the FLRW geometry. First, since $P_{-1/2+i\omega}(1)=1=J_0(0)$, we can see 
that~\eqref{2pcfB3} agrees with~\eqref{2pcfFL} when $\Delta\rho=\Delta\rho_0=0=\Delta z$, as they should. The main differences between these expressions are: (i) the law of cosines, equation~\eqref{deltarhoFL}, 
which now becomes the corresponding law in a two-dimensional hyperbolic space, 
namely~\eqref{deltarhoB3}, and (ii) the Bessel function in the kernel of the integral~\eqref{2pcfFL}, which is now replaced by a Legendre function. As we will see, all 
expressions agree at first order in the limit where $\Delta\eta\ll1$ in units of curvature scale 
(and hence, strictly speaking, even beyond the limit of coincident points that we commented above).

Apart from our physical requirements of coincident points, expression~\eqref{2pcfB3} is still quite 
general, inasmuch as it provides the form of the correlation between any two scalar functions in a 
space with $\mathbb{H}^2\times\mathbb{R}$ topology. However, we are only interested in situations in 
which this 2pcf does not lead to large modifications in the CMB data from the results of the flat 
FLRW case, since we know from observations that our universe is well fitted by those data. With this 
motivation, we can think of the CMB observations as taking place in a region described  just by a 
small patch of a larger BIII region. If that is the case, small curvature corrections from the global BIII metric might be lurking beyond the particle horizon scale, and their signatures should be encoded in small corrections to the large-angle CMB spectrum. Such corrections come from two different terms. The first of them is the function $P_{-\frac{1}{2}+i\omega}(\cosh\Delta\rho)$, which can be written as a powers series in 
$\Delta\rho$ (see Appendix~\ref{approxPJ}):
\be\label{approxPB3}
P_{-\frac{1}{2}+i\omega}(\cosh\Delta\rho) = J_0(\omega\Delta\rho)-\frac{(\Delta\rho)^2}{12}
\left[J_0(\omega\Delta\rho)-\frac{J_1(\omega\Delta\rho)}{2\omega\Delta\rho}\right]+\cdots
\ee
The second correction comes from the law of cosines~\eqref{deltarhoB3}, which can be expanded in 
powers of $\rho$ and $\rho'$ as
\be
\label{deltarhoB3approx}
(\Delta\rho)^2 = (\Delta\rho_0)^2 + \frac{\sin^2\Delta\varphi}{3}(\rho\rho')^2+\cdots
\ee
where $\Delta\rho_0$ was defined in equation~\eqref{deltarhoFL}. Both expansions above are exact, 
and can be carried at any desired order in their perturbation parameters. In our case, this 
parameter is the distance to the last scattering surface \emph{in units of the curvature scale}, 
$\Delta\eta$, which relates to $\Delta\rho$, $\rho$ and $\rho'$ through 
equations~\eqref{var2},~\eqref{deltarhoFL} and~\eqref{deltarhoB3approx}. Notice that this is the 
only free parameter appearing in our equations. If we thus combine equations~\eqref{approxPB3} 
and~\eqref{deltarhoB3approx}, retaining only corrections of order $(\Delta\eta)^2$, we arrive at
\be
P_{-\frac{1}{2}+i\omega}(\cosh\Delta\rho) \approx J_0(x) + 
\frac{(\rho\rho')^2\sin^2\Delta\varphi}{6(\Delta\rho_0)^2}\left[x\frac{dJ_0(x)}{dx}\right]
-\frac{(\Delta\rho_0)^2}{12}\left[J_0(x)-\frac{J_1(x)}{2x}\right]\,,
\ee
where $x=\omega\Delta\rho_0$. Inserting this result in equation~\eqref{2pcfB3} we finally have that
\begin{align}
\label{mainres1}
C(\hat{\mathbf{n}},\hat{\mathbf{n}}') \approx\; & \frac{1}{(6\pi)^2}\int_0^\infty\omega\dd\omega
\int_{-\infty}^{+\infty}\dd k\, {\cal P}(\tilde{q})J_0(\omega\Delta\rho_0)e^{ik\Delta z}\nn\\
& + \frac{1}{(6\pi)^2}\left[\frac{(\rho\rho')^2\sin^2\Delta\varphi}{6(\Delta\rho_0)^2 }
\int_0^\infty\omega\dd\omega\int_{-\infty}^{+\infty}\dd k\, {\cal 
P}(\tilde{q})\,(\omega\Delta\rho_0) 
\frac{dJ_0(\omega\Delta\rho_0)}{d(\omega\Delta\rho_0)}e^{ik\Delta z}\right.\nn\\
& \qquad\left.  - \frac{(\Delta\rho_0)^2}{12}\int_0^\infty\omega\dd\omega\int_{-\infty}^{
+\infty } \dd k\, {\cal P}(\tilde{q})\,\left[J_0(\omega\Delta\rho_0) - 
\frac{J_1(\omega\Delta\rho_0)}{2\omega\Delta\rho_0}\right]e^{ik\Delta z}\right]\,.
\end{align}

Expression~\eqref{mainres1} is one of our main results. Note that the effect of the 
anisotropic curvature will appear both at the isotropic ($\hat{\mathbf{n}}=\hat{\mathbf{n}}'$) and 
anisotropic ($\hat{\mathbf{n}}\neq\hat{\mathbf{n}}'$) parts of the CMB spectrum. We will investigate 
these signatures with more details in Section \ref{numerics}.

\subsection{Kantowski-Sachs case}

The computation of the 2pcf in the KS case follows the same lines as for BIII. Starting 
from~\eqref{temp2pcf}, we decompose the gravitational potential using the 
eigenfunctions~\eqref{KSeigenfunc}. Since we are considering a homogeneous but spatially 
anisotropic geometry, the correlation between modes must take the form
\be
\label{fourier2pcfKS}
\langle\Phi_{\ell m}(k)\Phi^{*}_{\ell' m'}(k')\rangle = {\cal 
\bar{P}}(\mathbf{q})\delta_{\ell\ell'}\delta_{mm'}\delta(k-k')\,,
\ee
where $\mathbf{q}\equiv(\ell,m,k)$. Using arguments similar to those given in the previous 
section, one can show that
\be\label{peksc}
{\cal \bar{P}}(\mathbf{q}) = 2\pi{\cal P}(\bar{q}).
\ee
Here we have defined $ \bar{q}^2\equiv k^2+(\ell+1/2)^2=q^2+1/4$ to 
account for the difference between the FLRW dispersion relation and \eqref{dispersionKS}, and we 
have chosen the numerical factor in the power spectrum so as to formally reproduce the expression 
for the 2pcf of the flat FLRW case in the limit of coincident points. Using the above power spectrum 
and the summation theorem~\eqref{legendre_id3} we find
\be
\label{2pcfKS}
C(\hat{\mathbf{n}},\hat{\mathbf{n}}') = \frac{1}{(6\pi)^2}\sum_{\ell=1}^\infty \left(\ell + 
\frac{1}{2}\right) \int_{-\infty}^{+\infty} \dd k\, {\cal P}(\bar{q})P_{\ell}(\cos\Delta\rho)
e^{ik\Delta z},
\ee
where
\be
\label{deltarhoKS}
\cos\Delta\rho \equiv \cos\rho\cos\rho'+\sin\rho\sin\rho'\cos(\varphi-\varphi)\,.
\ee
A remark is in order, concerning the derivation of~\eqref{peksc} and~\eqref{2pcfKS}. In the limit 
of coincident points, equation~\eqref{2pcfFL} becomes $(6\pi)^{-2}\int\omega\dd\omega\int\dd k\, {\cal 
P}(q)$. Clearly, the integral over $\omega$ in this expression can be split as a sum of all  the 
integrals over two consecutive positive integers. Assuming that the function ${\cal P}$ does not 
vary much with $\omega$ in any such interval, we can evaluate it at its middle point and then, 
in the considered case of coincident points, we are simply left with the integrals of the 
remaining factor $\omega\dd \omega$. In the interval 
$[\ell,\ell+1]$, e.g., this integral is equal to $(\ell+1/2)$. Hence, with our assumption, the 
isotropic 2pcf becomes  $(6\pi)^{-2}\sum_{\ell}(\ell+1/2)\int\dd k\, {\cal P}(\bar{q})$. The 
requirement 
that the 2pcf in the KS space should give an equivalent result in the limit of coincident points 
then justify~\eqref{peksc}. As a final comment, note that the main difference of the 2pcf in the KS 
space, apart from the modifications in the dispersion relation and the eigenfunctions, is the 
appearance of a sum over $\ell$ instead of an integral over $\omega$. This is reminiscent of the 
fact that the KS geometries have closed spatial subspaces in their spatial sections, leading to 
discrete eigenvalues of the Laplace-Beltrami operator.

Moving forward, we follow the same line of reasoning as in the previous section and consider the 
relevant region for the CMB observations as a small patch of a larger KS space. Consequently, we 
expand $P_\ell(\cos\Delta\rho)$ as a power series in $\Delta\rho$ [see~\eqref{PJexpansionKS} in the 
Appendix] to find
\be
P_\ell(\cos\Delta\rho) = J_0(\tilde{\ell}\Delta\rho)+\frac{(\Delta\rho)^2}{12}
\left[J_0(\tilde{\ell}\Delta\rho)-\frac{J_1(\tilde{\ell}\Delta\rho)}{2\tilde{\ell}
\Delta\rho}\right]+\cdots
\ee
where we have used the notation $\tilde{\ell}=\ell+1/2$. Note that, in contrast 
to~\eqref{approxPB3}, the first non-trivial correction now appears with a positive sign. 
Likewise, we expand $(\Delta\rho)^2$ defined in~\eqref{deltarhoKS} as a power series, which gives
\be
(\Delta\rho)^2 = (\Delta\rho_0)^2-\frac{\sin^2\Delta\varphi}{3}(\rho\rho')^2+\cdots
\ee
Combining these two expressions and retaining powers up to $(\Delta\eta)^2$, as before, we 
find
\be
P_\ell(\cos\Delta\rho)\approx J_0(x) - \frac{(\rho\rho')^2\sin^2\Delta\varphi}{6(\Delta\rho_0)^2}
\left[x\frac{dJ_0(x)}{dx}\right]+\frac{(\Delta\rho_0)^2}{12}\left[J_0(x)-\frac{J_1(x)}{2x}\right]\,,
\ee
where $x=(\ell+1/2)\Delta\rho_0$. It is worth noticing that, since we are assuming that 
$\Delta\rho_0\propto\Delta\eta$ is small, the Bessel functions in these expressions do not vary much 
in each interval $[\ell,\ell+1]$. Thanks to this fact, our discussion below \eqref{deltarhoKS} still 
applies and allows us to regard the sum over $\ell$ as a discretized approximation of the integral 
over $\dd\omega$. Combining all these results, we finally find
\begin{align}
\label{mainres2}
C(\hat{\mathbf{n}},\hat{\mathbf{n}}') \approx\; & \frac{1}{(6\pi)^2}\int_0^\infty\omega\dd\omega
\int_{-\infty}^{+\infty}\dd k\, {\cal P}(\bar{q})J_0(\omega\Delta\rho_0)e^{ik\Delta z}\nn\\
& - 
\frac{1}{(6\pi)^2}\left[\frac{(\rho\rho')^2\sin^2\Delta\varphi}{6(\Delta\rho_0)^2}
\int_0^\infty\omega\dd\omega\int_{-\infty}^{+\infty}\dd k\, {\cal P}(\bar{q})\,(\omega\Delta\rho_0) 
\frac{dJ_0(\omega\Delta\rho_0)}{d(\omega\Delta\rho_0)}e^{ik\Delta z}\right.\nn\\
& \qquad \left. - \frac{(\Delta\rho_0)^2}{12}\int_0^\infty\omega\dd\omega\int_{-\infty}^{
+\infty } \dd k\, {\cal P}(\bar{q})\, \left[J_0(\omega\Delta\rho_0) - 
\frac{J_1(\omega\Delta\rho_0)}{2\omega\Delta\rho_0}\right]e^{ik\Delta z}\right]\,,
\end{align}
with $\bar{q}^2$ evaluated at $k^2+\omega^2$ in these integrals. This is another of our main 
results.

\section{Qualitative analysis}\label{numerics}

We will now turn to a numerical analysis in order to extract some qualitative information from 
formulas~\eqref{mainres1} and~\eqref{mainres2}. We start by noticing that these expressions 
can be summarized in a single formula as
\be
C(\hat{\mathbf{n}},\hat{\mathbf{n}}') = C(\vartheta) \pm {\cal 
F}(\hat{\mathbf{n}},\hat{\mathbf{n}}')\,,
\ee
where $C(\vartheta)$, with $\vartheta=\arccos(\hat{\mathbf{n}}\cdot\hat{\mathbf{n}}')$, is the 
isotropic 2pcf and where the anisotropic function ${\cal F}(\hat{\mathbf{n}},\hat{\mathbf{n}}')$ is 
defined by the additional terms in the last two lines of equation~\eqref{mainres1} or 
equation~\eqref{mainres2}. Here, the plus and minus signs correspond to the BIII and KS cases, 
respectively. Alternatively, our results can be written in terms of the temperature \emph{covariance 
matrix}, defined in terms of the temperature 
harmonic coefficients, $a_{\ell m}=\int\dd^2\hat{\mathbf{n}}\Delta 
T(\hat{\mathbf{n}})Y^*_{\ell m}(\hat{\mathbf{n}})$, 
as
\be
\langle a_{\ell m}a^*_{\ell' m'}\rangle = 
\int\dd^2\hat{\mathbf{n}}\int\dd^2\hat{\mathbf{n}}'\,C(\hat{\mathbf{n}},\hat{\mathbf{n}}')\,
Y^*_{\ell m}(\hat{\mathbf{n}})Y_{\ell'm'}(\hat{\mathbf{n}}')\,.
\ee
Decomposing each term in $C(\hat{\mathbf{n}},\hat{\mathbf{n}}')$ into spherical harmonics, we find
\be
\langle a_{\ell m}a^*_{\ell' m'}\rangle = C_\ell\delta_{\ell\ell'}\delta_{mm'}
\pm\langle{\cal F}_{\ell m\ell'm'}\rangle\,.
\ee

\begin{figure}
\begin{centering}
\includegraphics[scale=0.6]{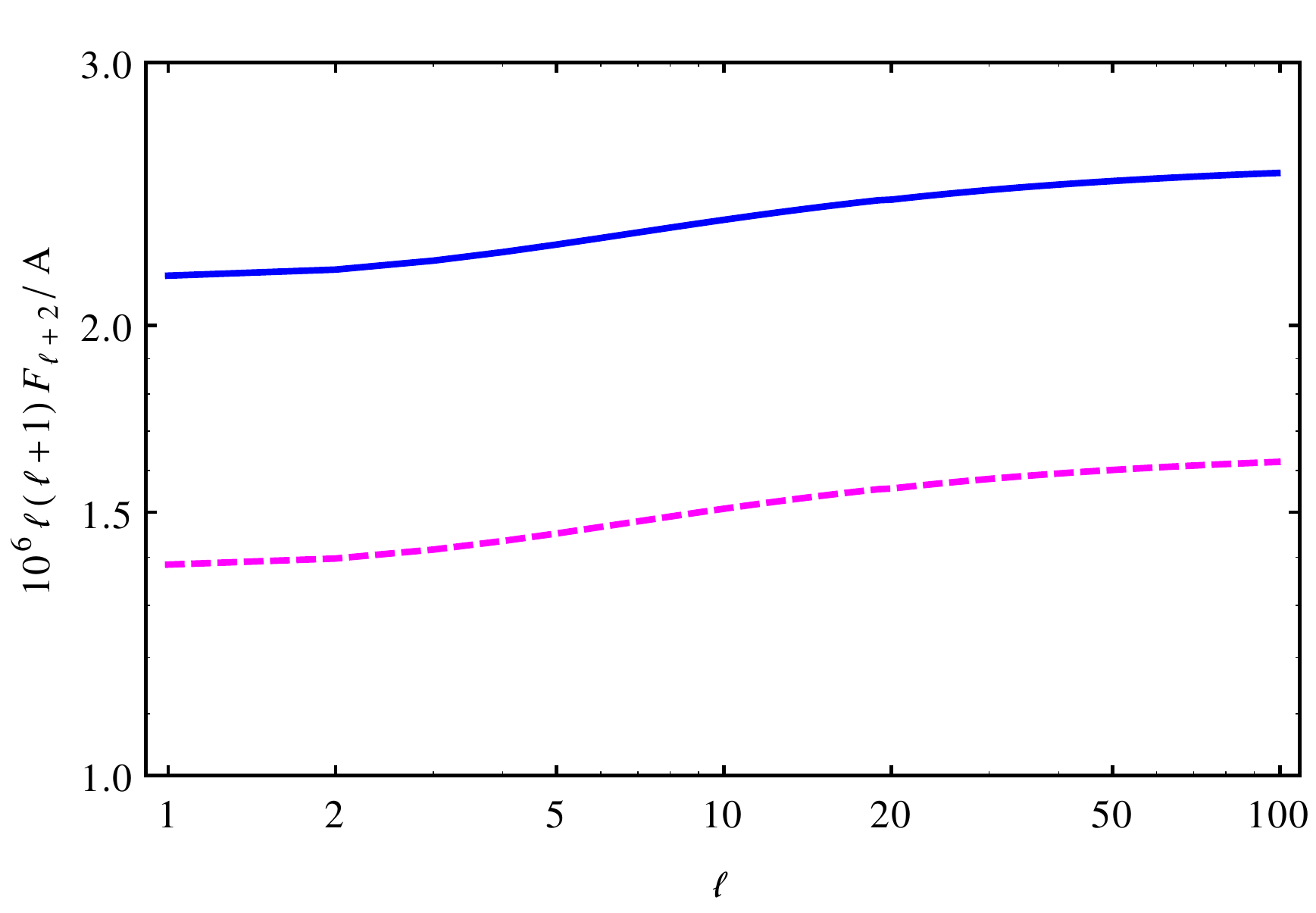}
\par\end{centering}
\caption{Log-log plot of the quantity $\ell(\ell+1){\cal F}_{\ell+2}$ as a function of the angular multipole $\ell$, and in units of the power spectrum amplitude $A$. The curves show the results of our numerical computations assuming $\Delta\eta$ equal to 0.04 (dashed, magenta) and 0.05 
(continuous, blue), in units of the curvature scale.\label{figure1}}
\end{figure}

There are two types of corrections coming from the anisotropy of the spatial curvature. The first of 
them arises from the diagonal part of the matrix $\langle{\cal F}_{\ell m\ell'm'}\rangle$. Such 
correction leads to an effective temperature power spectrum which differs from $C_\ell$ mainly at 
small multipoles. Unfortunately this effect does not offer a strong constraint on the model since, 
besides being strongly limited by cosmic variance, it is degenerate with isotropic physical 
mechanisms which also alter the low-$\ell$ tail of the spectrum (e.g. the Integrated Sachs-Wolfe 
effect). The second type of correction,on the other hand, arises from off-diagonal terms in the 
covariance matrix, and is a genuine signature of statistical anisotropy. Moreover, this correction 
is present at all multipolar scales, a feature which alleviates the problem of cosmic variance. The 
simplest off-diagonal term we could expect to find is of the form $\langle a_{\ell 
m}a^*_{(\ell+1)m}\rangle=\langle F_{\ell m(\ell+1) m}\rangle$. Since this term couples 
even and odd multipoles, it is a signature of parity breaking. However, these terms are exactly zero 
in our case, since the metrics we are considering are invariant under the parity transformation 
$(\rho,\varphi,z)\rightarrow(\rho,\varphi\pm\pi,-z)$. The next off-diagonal term we can calculate 
is $\langle a_{\ell m}a^*_{(\ell+2)m}\rangle$, where $-\ell\leq m \leq \ell$. So, for each
$\ell$ there will $2\ell+1$ terms of this type. Recall also that $a_{\ell m}=(-1)^m a^*_{\ell,-m}$, 
as required by the reality of $\Delta T(\hat{\mathbf{n}})$. Then, to further reduce the effect 
of cosmic variance, we will work with an azimuthally averaged measure of anisotropy defined as a sum 
over the $2\ell+1$ possible realizations of the label $m$. Therefore, we define the following (real) 
quantity:
\be
{\cal F}_{\ell+2}\equiv \frac{1}{2\ell+1}\sum_{m=-\ell}^\ell |\langle a_{\ell 
m}a^*_{(\ell+2)m}\rangle|\,.
\ee
We have evaluated this quantity numerically using a power spectrum per logarithmic band of the 
Harrison-Zel'dovich form: $q^3{\cal P}(q)=A=\mbox{constant}$, and two different choices of the 
parameter $\Delta\eta$ compatible with our hypothesis that $\Delta\eta\ll1$. We show in 
figure~\eqref{figure1} a plot of $\ell(\ell+1){\cal F}_{\ell+2}$ for a wide range of the multipole 
$\ell$. Our analysis shows that the spectrum grows smoothly with $\ell$ up to $\ell=100$, where the 
use of the Sachs-Wolfe effect alone is justified. 

\section{Conclusion}\label{conclusions}

The hypothesis that we live in a spatially infinite, homogeneous, and isotropic universe is 
one of the central pillars of modern cosmology. Besides being strongly supported by CMB
measurements, the simplifications introduced by this hypothesis in the framework of general 
relativity has allowed cosmologists to build a robust description of the large-scale universe 
with only six free parameters. Nonetheless, the fact that computations are simpler in the 
framework of an FLRW metric should not prevent us from considering more general geometries, 
especially when one realizes that the observed isotropy of the CMB does not rule out less 
symmetric models of the universe. Indeed, it is known that the anisotropies of the background metric 
can manifest themselves through other channels different from the simple anisotropic expansion 
of the scale factor. 

In this work we have explored the ideas introduced in references 
~\cite{Mimoso:1993ym,Carneiro:2001fz,Koivisto:2010dr}, in which the anisotropy of the universe does 
not result from its dynamical expansion, but rather from the curvature of its 
spatial sections. In the context of three-dimensional spaces, this feature can be implemented 
naturally only in two different ways: either by considering the Bianchi type III or the 
Kantowski-Sachs metric. In both models the expansion of the universe is controlled by a single 
scale factor, and hence the resulting CMB field is exactly isotropic at the background 
level~\cite{Carneiro:2001fz}. Moreover, the fact that the dynamics are dictated by a single 
scale factor introduces simplifications which render the model tractable from a 
computational point of view. Based on this property, and starting from the full theory of linear and 
gauge-invariant perturbations introduced in~\cite{pcm}, we have established several important 
results. First, by means of a simple parameterization of the considered metrics in cylindrical 
coordinates, we have given explicit 
expressions for a complete set of spatial eigenfunctions possessing the symmetries of these 
spaces. This step is crucial for the comparison of the theory with observations and, to the 
best of our knowledge, has not appeared before in the context of anisotropic cosmologies. Second, we 
have shown that, provided that the early universe has expanded in accordance with the inflationary 
paradigm, large-scale and adiabatic scalar perturbations will be described by one single 
gravitational potential, $\Phi$, just as in standard perturbation theory in isotropic universes. 
Since the main large-angle CMB effects depend only on the photon geodesics in the perturbed 
spacetime, our result shows that the temperature fluctuations at these scales will have exactly the 
same functional dependence on $\Phi$ as the one found in the isotropic setup. In particular, the 
Sachs-Wolfe effect in both BIII and KS spaces is given by $\Delta T = \Phi/3$.

Focusing on the Sachs-Wolfe effect, which is the most important effect for CMB at large angles, and 
with the help of the basis eigenfunctions that we have found, we have explicitly computed the 
spatial two-point correlation function in the BIII and KS geometries in the limit of small 
$\Delta\eta$, that is, a small ratio of the distance to the last scattering surface to the curvature 
radius. There are two different corrections to the isotropic correlation function, one resulting 
from the expansion of the spatial eigenfunctions around their isotropic counterpart, and 
another resulting from the law of cosines in curved spaces. The general effect of these corrections 
is two-fold: first, it changes the low-$\ell$ tail of the isotropic spectrum, and second, it 
induces off-diagonal temperature correlations which ought not to be present if the 
universe is isotropic. We have evaluated the latter numerically, and we have found that the 
effect goes beyond the low multipole region of the spectrum, corresponding to small $\ell$, 
region where, at least in principle, these correlations could be detected. It is important to note 
that we cannot predict the ratio of the last scattering surface to the curvature radius (i.e., 
$\Delta\eta$), which in this work was treated as a small free parameter. Nonetheless, current 
bounds on superhorizon features are usually of the order of a few percent~\cite{Erickcek:2008jp} or
even less~\cite{Castro:2003bk}, which means that our choices for $\Delta\eta$ are based on
educated guesses.

Finally, it is worth noticing that the formalism that we have developed here is quite general 
and can be directly applied to situations which include other effects, such as the integrated 
Sachs-Wolfe and Doppler effects, as well as in scenarios involving more general matter fields. This 
work also paves the way to more challenging investigations, such as the effect of anisotropic 
curvature in the spectrum of gravitational waves and the polarization of the CMB. Such 
analyses, together with a thorough assessment of the detectability of these effects with actual 
and upcoming experimental probes, are currently in progress.

\paragraph*{Note Added} After this work was submitted to publication we became aware of 
references~\cite{Adamek:2010sg,BlancoPillado:2010uw,Graham:2010hh}, in which the spatial eigenfunctions of B3 and KS spacetimes were found independently. We thank Julian Adamek for bringing these references to our attention.

\acknowledgments
We would like to thank Cyril Pitrou and Mario Cesar Baldiotti for useful discussion during the 
preparation of this work. S.C. and T.S.P thank the Brazilian agency CNPq for financial support. 
T.S.P. thanks the Instituto de Estrutura de la Materia in Spain, IEM-CSIC, for its hospitality 
during the realization of this work. G.A.M.M. acknowledges financial support from the research 
MICINN/MINECO Projects No. FIS2011-30145-C03-02 and FIS2014-54800-C2-2-P from Spain.

\appendix

\section{Mathematical complements}
In this appendix we succinctly review the main mathematical tools used in this work. 

\subsection{Summation theorems}

The derivation of the 2pcf in flat (FLRW) geometries makes use of the following identity
\be
\label{bessel_id1}
\sum_{m=-\infty}^{+\infty} J_m(\omega\rho)J_m(\omega\rho')e^{im(\varphi-\varphi')} 
= J_0(\omega\Delta\rho_0)\,,
\ee
where $\Delta\rho_0$ is defined in \eqref{deltarhoFL}. The derivation of the 2pcf in the open 
(BIII) case, on the other hand, relies on the identities
\begin{align}
\label{legendre_id1}
P_{\nu}^{-m}(z) & =\frac{\Gamma(\nu-m+1)}{\Gamma(\nu+m+1)}P_{\nu}^{m}(z)\,,\\ \label{legendre_id1b}
P_{-\frac{1}{2}+i\omega}(\cosh\Delta\rho) & =\sum_{m=-\infty}^{+\infty}(-1)^{m}P_{-\frac{1}{2}+i\omega}^{-m}(\cosh\rho)P_{-\frac{1}{2}+i\omega}^{m}(\cosh\rho')e^{im(\varphi-\varphi')}\,,
\end{align}
where $\Delta\rho$ is defined in~\eqref{deltarhoB3}. Analogously, the derivation of the 2pcf in the 
closed case uses the following formulas:
\begin{align}
\label{legendre_id2}
P_{\ell}^{-m}(x) & =(-1)^m\frac{(\ell-m)!}{(\ell+m)!}P_{\ell}^{m}(x)\,,\\
\label{legendre_id3}
P_{\ell}(\cos\Delta\rho) & =\sum_{m=-\infty}^{+\infty}(-1)^{m}P_{\ell}^{-m}(\cos\rho)P_{\ell}^{m}(\cos\rho')e^{im(\varphi-\varphi')}\,,
\end{align}
where $\Delta\rho$ is defined now in~\eqref{deltarhoKS}.

\subsection{Approximating Legendre polynomials by Bessel functions}
\label{approxPJ}

In order to find an approximation to $P_{-1/2+i\omega}(\cosh y)$ in terms of Bessel functions we 
start with the following integral representations~\cite{jeffrey2007table}:
\begin{align}
\label{intrep1}
P_{-\frac{1}{2}+i\omega}(\cosh y) & =\frac{\sqrt{2}}{\pi}\int_{0}^{y}\frac{\cos\omega t}{\sqrt{\cosh 
y-\cosh t}}\, \dd t\,,\qquad y\geq0\,,\\
J_{n}(x) & =\frac{\left(x/2\right)^{n}}{\Gamma(n+1/2)\Gamma(1/2)}\int_{-1}^{1}\left(1-t^{2}\right)^{n-1/2}\cos xt\, \dd t\,.
\end{align}
We now define $t=uy$, $x=\omega y$, and write the argument of integral~\eqref{intrep1} as a power 
series in $y$. This gives
\begin{align}
P_{-\frac{1}{2}+i\omega}\left(\cosh y\right) & =\frac{\sqrt{2}}{\pi}\int_{0}^{1}\frac{y \,\cos xu}
{\sqrt{\cosh y -\cosh uy}}\dd u\nonumber \\
& =\frac{1}{\pi}\int_{0}^{1}\cos xu\,
\left[
\frac{2}{\sqrt{1-u^2}}-\frac{y^2(1+u^2)}{12\sqrt{1-u^2}}+\cdots\right]\dd u \nn \\
\label{PJexpansionB3}
& = J_{0}(x)-\frac{y^{2}}{12}\left[J_{0}(x)-\frac{J_{1}(x)}{2x}\right]+\cdots
\end{align}
Note that this series expansion is exact. In particular,  since the resulting series is controlled 
by $y$, we can truncate it at any desired order in this variable if we regard $x$ as a fixed 
number.

Following the same procedure as above, but now using
\[
P_\ell(\cos\theta) = \frac{\sqrt{2}}{\pi}\int_0^\theta\frac{\cos(\ell+1/2)t}{\sqrt{\cos t-\cos\theta}}\,\dd t\,,
\qquad 0\leq\theta\leq\pi\,,
\]
one can show that, defining $x=(\ell+1/2)\theta$, one has~\cite{jeffrey2007table}
\begin{align}
\label{PJexpansionKS}
P_{\ell}(\cos\theta) & = J_0(x) + 
\frac{\theta^2}{12}\left[J_0(x)-\frac{J_1(x)}{2x}\right]+\cdots
\end{align}
Again, note that the series is controlled by $\theta$, for constant $x$.

The reader familiar with the so-called flat sky approximation (usually employed in CMB analysis) 
will recognize expansion~\eqref{PJexpansionKS}. It says that, for small patches of the CMB sky, the 
eigenfunctions of the sphere, i.e., the Legendre functions, converge towards eigenfunctions of the 
plane, i.e, Bessel functions~\cite{Bernardeau:2010ac}. For the same reason, 
expression~\eqref{PJexpansionB3} gives the eigenfunctions of the pseudo-sphere in terms of 
eigenfunctions of the plane.

\bibliographystyle{JHEP}
\addcontentsline{toc}{section}{\refname}\bibliography{b3ks_signature}

\providecommand{\href}[2]{#2}\begingroup\raggedright\begin{thebibliography}{10}

\bibitem{GarciaBellido:2008nz}
J.~Garc\'ia-Bellido and T.~Haugboelle, {\it {Confronting Lemaitre-Tolman-Bondi
  models with observational cosmology}},  {\em JCAP} {\bf 0804} (2008) 003,
  [\href{http://arxiv.org/abs/0802.1523}{{\tt arXiv:0802.1523}}].

\bibitem{Nadathur:2010zm}
S.~Nadathur and S.~Sarkar, {\it {Reconciling the local void with the CMB}},
  {\em Phys.Rev.} {\bf D83} (2011) 063506,
  [\href{http://arxiv.org/abs/1012.3460}{{\tt arXiv:1012.3460}}].

\bibitem{Biswas:2010xm}
T.~Biswas, A.~Notari, and W.~Valkenburg, {\it {Testing the void against
  cosmological data: Fitting CMB, BAO, SN and H0}},  {\em JCAP} {\bf 1011}
  (2010) 030, [\href{http://arxiv.org/abs/1007.3065}{{\tt arXiv:1007.3065}}].

\bibitem{Yoo:2010qy}
C.-M. Yoo, K.-i. Nakao, and M.~Sasaki, {\it {CMB observations in LTB universes:
  Part I: Matching peak positions in the CMB spectrum}},  {\em JCAP} {\bf 1007}
  (2010) 012, [\href{http://arxiv.org/abs/1005.0048}{{\tt arXiv:1005.0048}}].

\bibitem{martinez1995delta}
E.~Mart\'inez-Gonz\'alez and J.~Sanz, {\it $\delta t/t$ and the isotropy of the
  universe},  {\em Astron.Astrophys.} {\bf 300} (1995) 346.

\bibitem{Maartens:1994qq}
R.~Maartens, G.~F. Ellis, and S.~Stoeger, William~R., {\it {Limits on
  anisotropy and inhomogeneity from the cosmic background radiation}},  {\em
  Phys.Rev.} {\bf D51} (1995) 1525,
  [\href{http://arxiv.org/abs/astro-ph/9501016}{{\tt astro-ph/9501016}}].

\bibitem{Maartens:1995hh}
R.~Maartens, G.~F. Ellis, and S.~Stoeger, William~R., {\it {Anisotropy and
  inhomogeneity of the universe from Delta(T) / T}},  {\em Astron.Astrophys.}
  {\bf 309} (1996) L7, [\href{http://arxiv.org/abs/astro-ph/9510126}{{\tt
  astro-ph/9510126}}].

\bibitem{Campanelli:2006vb}
L.~Campanelli, P.~Cea, and L.~Tedesco, {\it {Ellipsoidal universe can solve The
  CMB quadrupole problem}},  {\em Phys.Rev.Lett.} {\bf 97} (2006) 131302,
  [\href{http://arxiv.org/abs/astro-ph/0606266}{{\tt astro-ph/0606266}}].

\bibitem{Gumrukcuoglu:2006xj}
A.~G{\"u}mr{\"u}k{\c c}{\" u}o{\u g}lu, C.~R. Contaldi, and M.~Peloso, {\it
  {CMB anomalies from relic anisotropy}},
  \href{http://arxiv.org/abs/astro-ph/0608405}{{\tt astro-ph/0608405}}.

\bibitem{Pontzen:2007ii}
A.~Pontzen and A.~Challinor, {\it {Bianchi model CMB polarization and its
  implications for CMB anomalies}},  {\em Mon.Not.Roy.Astron.Soc.} {\bf 380}
  (2007) 1387, [\href{http://arxiv.org/abs/0706.2075}{{\tt arXiv:0706.2075}}].

\bibitem{Rodrigues:2007ny}
D.~C. Rodrigues, {\it {Anisotropic cosmological constant and the CMB quadrupole
  anomaly}},  {\em Phys.Rev.} {\bf D77} (2008) 023534,
  [\href{http://arxiv.org/abs/0708.1168}{{\tt arXiv:0708.1168}}].

\bibitem{Pitrou:2008gk}
C.~Pitrou, T.~S. Pereira, and J.-P. Uzan, {\it {Predictions from an anisotropic
  inflationary era}},  {\em JCAP} {\bf 0804} (2008) 004,
  [\href{http://arxiv.org/abs/0801.3596}{{\tt arXiv:0801.3596}}].

\bibitem{Koivisto:2008ig}
T.~Koivisto and D.~F. Mota, {\it {Anisotropic dark energy: Dynamics of
  background and perturbations}},  {\em JCAP} {\bf 0806} (2008) 018,
  [\href{http://arxiv.org/abs/0801.3676}{{\tt arXiv:0801.3676}}].

\bibitem{Pitrou:2015iya}
C.~Pitrou, T.~S. Pereira, and J.-P. Uzan, {\it {Weak-lensing by the large scale
  structure in a spatially anisotropic universe: theory and predictions}},
  {\em Phys. Rev.} {\bf D92} (2015), no.~2 023501,
  [\href{http://arxiv.org/abs/1503.01125}{{\tt arXiv:1503.01125}}].

\bibitem{Pereira:2015jya}
T.~S. Pereira, C.~Pitrou, and J.-P. Uzan, {\it {Weak-lensing $B$-modes as a
  probe of the isotropy of the universe}},
  \href{http://arxiv.org/abs/1503.01127}{{\tt arXiv:1503.01127}}.

\bibitem{Barrow:1997sy}
J.~D. Barrow, {\it {Cosmological limits on slightly skew stresses}},  {\em
  Phys. Rev.} {\bf D55} (1997) 7451--7460,
  [\href{http://arxiv.org/abs/gr-qc/9701038}{{\tt gr-qc/9701038}}].

\bibitem{Barrow:1997mj}
J.~D. Barrow, P.~G. Ferreira, and J.~Silk, {\it {Constraints on a primordial
  magnetic field}},  {\em Phys. Rev. Lett.} {\bf 78} (1997) 3610--3613,
  [\href{http://arxiv.org/abs/astro-ph/9701063}{{\tt astro-ph/9701063}}].

\bibitem{Barrow:1998ih}
J.~D. Barrow and R.~Maartens, {\it {Anisotropic stresses in inhomogeneous
  universes}},  {\em Phys. Rev.} {\bf D59} (1999) 043502,
  [\href{http://arxiv.org/abs/astro-ph/9808268}{{\tt astro-ph/9808268}}].

\bibitem{Barrow:2001pi}
J.~D. Barrow and R.~Maartens, {\it {Kaluza-Klein anisotropy in the CMB}},  {\em
  Phys. Lett.} {\bf B532} (2002) 153--158,
  [\href{http://arxiv.org/abs/gr-qc/0108073}{{\tt gr-qc/0108073}}].

\bibitem{Barrow:1985tda}
J.~D. Barrow, R.~Juszkiewicz, and D.~H. Sonoda, {\it {Universal rotation - How
  large can it be?}},  {\em Mon.Not.Roy.Astron.Soc.} {\bf 213} (1985) 917--943.

\bibitem{Mimoso:1993ym}
J.~P. Mimoso and P.~Crawford, {\it {Shear-free anisotropic cosmological
  models}},  {\em Classical Quant. Grav.} {\bf 10} (1993) 315.

\bibitem{cmm}
S.~Carneiro and G.~A. Mena~Marug{\'a}n, {\it {An anisotropic cosmological model
  with isotropic background radiation}},  {\em Lect. Notes Phys.} {\bf 617}
  (2003) 302, [\href{http://arxiv.org/abs/gr-qc/0203025}{{\tt gr-qc/0203025}}].

\bibitem{Carneiro:2001fz}
S.~Carneiro and G.~A. Mena~Marug{\'a}n, {\it {Anisotropic cosmologies
  containing isotropic background radiation}},  {\em Phys.Rev.} {\bf D64}
  (2001) 083502, [\href{http://arxiv.org/abs/gr-qc/0109039}{{\tt
  gr-qc/0109039}}].

\bibitem{Koivisto:2010dr}
T.~S. Koivisto, D.~F. Mota, M.~Quartin, and T.~G. Zlosnik, {\it {On the
  possibility of anisotropic curvature in cosmology}},  {\em Phys.Rev.} {\bf
  D83} (2011) 023509, [\href{http://arxiv.org/abs/1006.3321}{{\tt
  arXiv:1006.3321}}].

\bibitem{Menezes:2012kc}
R.~S. Menezes~Jr., C.~Pigozzo, and S.~Carneiro, {\it {Distance-redshift
  relations in an anisotropic cosmological model}},  {\em JCAP} {\bf 1303}
  (2013) 033, [\href{http://arxiv.org/abs/1210.2909}{{\tt arXiv:1210.2909}}].

\bibitem{Miranda:2014ema}
W.~Miranda, S.~Carneiro, and C.~Pigozzo, {\it {SNe Ia Tests of quintessence
  tracker cosmology in an anisotropic background}},  {\em JCAP} {\bf 1407}
  (2014) 043, [\href{http://arxiv.org/abs/1405.3673}{{\tt arXiv:1405.3673}}].

\bibitem{pcm}
T.~S. Pereira, S.~Carneiro, and G.~A. Mena~Marug{\'a}n, {\it {Inflationary
  perturbations in anisotropic, shear-free universes}},  {\em JCAP} {\bf 1205}
  (2012) 040, [\href{http://arxiv.org/abs/1203.2072}{{\tt arXiv:1203.2072}}].

\bibitem{Lyth:1995cw}
D.~H. Lyth and A.~Woszczyna, {\it {Large scale perturbations in the open
  universe}},  {\em Phys.Rev.} {\bf D52} (1995) 3338,
  [\href{http://arxiv.org/abs/astro-ph/9501044}{{\tt astro-ph/9501044}}].

\bibitem{Bander:1965im}
M.~Bander and C.~Itzykson, {\it {Group theory and the Hydrogen atom. II}},
  {\em Rev. Mod. Phys.} {\bf 38} (1966) 346.

\bibitem{Peter:1208401}
P.~Peter and J.-P. Uzan, {\em {Primordial Cosmology }}.
\newblock Oxford Graduate Texts. Oxford Univ. Press, Oxford, 2009.

\bibitem{GS}
M.~H. Gerlach and U.~K. Sengupta, {\it {Gauge-invariant perturbations on most
  general spherically symmetric space-times}},  {\em Phys.Rev.} {\bf D19}
  (1979) 2268.

\bibitem{Pereira:2007yy}
T.~S. Pereira, C.~Pitrou, and J.-P. Uzan, {\it {Theory of cosmological
  perturbations in an anisotropic universe}},  {\em JCAP} {\bf 0709} (2007)
  006, [\href{http://arxiv.org/abs/0707.0736}{{\tt arXiv:0707.0736}}].

\bibitem{MenezesCarneiro}
R.~S. Menezes~Jr. and S.~Carneiro, {\em Private communication}.

\bibitem{Mukhanov:1990me}
V.~F. Mukhanov, H.~A. Feldman, and R.~H. Brandenberger, {\it {Theory of
  cosmological perturbations. Part 1. Classical perturbations. Part 2. Quantum
  theory of perturbations. Part 3. Extensions}},  {\em Phys. Rept.} {\bf 215}
  (1992) 203.

\bibitem{Abramo:2010gk}
L.~R. Abramo and T.~S. Pereira, {\it {Testing gaussianity, homogeneity and
  isotropy with the cosmic microwave background}},  {\em Adv. Astron.} {\bf
  2010} (2010) 378203, [\href{http://arxiv.org/abs/1002.3173}{{\tt
  arXiv:1002.3173}}].

\bibitem{Erickcek:2008jp}
A.~L. Erickcek, S.~M. Carroll, and M.~Kamionkowski, {\it {Superhorizon
  perturbations and the cosmic microwave background}},  {\em Phys.Rev.} {\bf
  D78} (2008) 083012, [\href{http://arxiv.org/abs/0808.1570}{{\tt
  arXiv:0808.1570}}].

\bibitem{Castro:2003bk}
P.~G. Castro, M.~Douspis, and P.~G. Ferreira, {\it {Scale of homogeneity of the
  universe from WMAP}},  {\em Phys.Rev.} {\bf D68} (2003) 127301,
  [\href{http://arxiv.org/abs/astro-ph/0309320}{{\tt astro-ph/0309320}}].

\bibitem{Adamek:2010sg}
J.~Adamek, D.~Campo, and J.~C. Niemeyer, {\it {Anisotropic Kantowski-Sachs
  Universe from Gravitational Tunneling and its Observational Signatures}},
  {\em Phys. Rev.} {\bf D82} (2010) 086006,
  [\href{http://arxiv.org/abs/1003.3204}{{\tt arXiv:1003.3204}}].

\bibitem{BlancoPillado:2010uw}
J.~J. Blanco-Pillado and M.~P. Salem, {\it {Observable effects of anisotropic
  bubble nucleation}},  {\em JCAP} {\bf 1007} (2010) 007,
  [\href{http://arxiv.org/abs/1003.0663}{{\tt arXiv:1003.0663}}].

\bibitem{Graham:2010hh}
P.~W. Graham, R.~Harnik, and S.~Rajendran, {\it {Observing the Dimensionality
  of Our Parent Vacuum}},  {\em Phys. Rev.} {\bf D82} (2010) 063524,
  [\href{http://arxiv.org/abs/1003.0236}{{\tt arXiv:1003.0236}}].

\bibitem{jeffrey2007table}
A.~Jeffrey and D.~Zwillinger, {\em Table of Integrals, Series, and Products}.
\newblock Academic Press, 2007.

\bibitem{Bernardeau:2010ac}
F.~Bernardeau, C.~Pitrou, and J.-P. Uzan, {\it {CMB spectra and bispectra
  calculations: Making the flat-sky approximation rigorous}},  {\em JCAP} {\bf
  1102} (2011) 015, [\href{http://arxiv.org/abs/1012.2652}{{\tt
  arXiv:1012.2652}}].

\end{thebibliography}\endgroup

\end{document}